\documentclass[12pt]{article}
\usepackage{latexsym}

\input epsf.sty
\newdimen\psfigsize
\def\psfigure#1 #2 #3 #4 #5{
    \begin{figure}[tbp]
    \vbox{
    \null\hskip#2\epsfxsize=#1 \epsfbox[0 0 4096 4096]{#4}
    \vskip 10truept
    \caption {#5 \label{#3}}
    \vskip 0.1truein plus0.2truein}
    \end{figure}
}
\def\pspagefigure#1 #2 #3 #4 #5{
    \begin{figure}[p]
    \vbox{
    \null\hskip#2\epsfxsize=#1 \epsfbox[0 0 4096 4096]{#4}
    \vskip 10truept
    \caption {#5 \label{#3}}
    \vskip 0.1truein plus0.2truein}
    \end{figure}
}
\def\psoddfigure#1 #2 #3 #4 #5 #6{
    \begin{figure}[tbhp]
    \vbox{
    \null\hskip#3\epsfxsize=#1 \epsfbox[0 0 4096 4096]{#5}
    \vskip -#1 \vskip #2 \vskip 10truept
    \vskip 10truept
    \caption {#6 \label{#4}}
    \vskip 0.1truein plus0.2truein}
    \end{figure}
}
\def\figurespace#1 #2 #3 #4 {
    \begin{figure}[tbhp]
    \vbox{
    \psfigsize=#1truein
    \vskip \psfigsize
    \vskip 10truept
    \caption {#4 \label{#3}}
    \vskip 0.1truein plus0.2truein}
    \end{figure}
}
\def\gnufigure#1 #2 #3 #4 #5 #6{
    \begin{figure}[tbhp]
    \vbox{
    \null\hskip#3\epsfxsize=#1 \epsfbox{#5}
    \vskip -#1 \vskip #2 \vskip 10truept
    \vskip 10truept
    \hbox{\null\hskip 1.0in \parbox[t]{4.5in}{ \caption {#6 \label{#4}} } }
    \vskip 0.1truein plus0.2truein}
    \end{figure}
}

\def\pbp{\bar\psi\psi}
\def\epbp{\langle\bar\psi\psi\rangle}

\newcommand{\eplaq}{\langle\Box\rangle}

\def\Tr{\mathop{\rm Tr}}

\def\Dslash{\mathop{\not\!\! D}}

\def\LL{\left\langle}	
\def\RR{\right\rangle}	
\def\LP{\left(}		
\def\RP{\right)}	
\def\PAR#1#2{ {{\partial #1}\over{\partial #2}} }
\def\PARTWO#1#2{ {{\partial^2 #1}\over{\partial #2}^2} }
\def\PARTWOMIX#1#2#3{ {{\partial^2 #1}\over{\partial #2 \partial #3}} }

\def\BE{\begin{equation}}
\def\EE{\end{equation}}
\def\BEA{\begin{eqnarray}}
\def\EEA{\end{eqnarray}}
\def\EL{\nonumber\\}
  
\newcommand{\gbeta}{6/g^2}
\newcommand{\la}[1]{\label{#1}}
\newcommand{\ie}{{\em i.e.\ }}
\newcommand{\eg}{{\em e.\,g.}}

\newcommand{\dt}{\Delta t}
 
\begin{document}
\title{\Large \bf The Equation of state for two flavor QCD at $\bf N_t=6$}
\author{{\bf Claude~Bernard} \\
{\small \it Department of Physics, Washington University, St.~Louis, MO 63130, USA }\\
{\bf Tom~Blum }\\
{\small \it Department of Physics, Brookhaven National Lab, Upton, NY 11973, USA }\\
{\bf Carleton~DeTar}\\
{\small \it Physics Department, University of Utah, Salt Lake City, UT 84112, USA }\\
{\bf Steven~Gottlieb, Kari~Rummukainen}
\thanks{ new address: Fakult\"at f\"ur Physik, Universit\"at Bielefeld,
D-33615, Bielefeld, Germany }\\
{\small \it Department of Physics, Indiana University, Bloomington, IN 47405, USA }\\
{\bf Urs~M.~Heller }\\
{\small \it SCRI, Florida State University, Tallahassee, FL 32306-4052, USA }\\
{\bf James~E.~Hetrick}\thanks{new address: Department of Physics,
Washington University, St.~Louis, MO 63130, USA}, {\bf Doug~Toussaint}\\
{\small \it Department of Physics, University of Arizona, Tucson, AZ 85721, USA }\\
{\bf Leo K\"arkk\"ainen}\\
{\small \it Nokia Research Center, P.O.\ Box 100, FIN-33721 Tampere, Finland }\\
{\bf Bob~Sugar }\\
{\small \it Department of Physics, University of California, Santa Barbara, CA 93106, USA }\\
{\bf Matthew~Wingate }\\
{\small \it Physics Department, University of Colorado, Boulder, CO 80309, USA }
} 

\maketitle

\begin{abstract}
We calculate the two flavor equation of state for QCD on lattices
with lattice spacing $a=(6T)^{-1}$ and find that cutoff effects are
substantially reduced compared to an earlier study using $a=(4T)^{-1}$.
However, it is likely that significant cutoff effects remain.
We fit the lattice data to expected forms of the free energy density
for a second order phase transition at zero-quark-mass, 
which allows us to extrapolate
the equation of state to $m_q=0$ and to extract the speed of sound. We
find that the equation of state depends weakly on the quark mass for small
quark mass.
\end{abstract}


\section{Introduction}

It is generally believed that at high temperatures QCD undergoes
either a phase transition or a fairly sharp crossover into a regime
where hadrons ``dissolve'' into a quark-gluon plasma.
Testing this scenario is a major goal of current and planned
experiments in heavy ion collisions.  Although the fireball created in
such a collision is at best in quasi-equilibrium, knowledge of the
equilibrium equation of state for QCD is nonetheless very useful for
constraining the parameters of models of the quark-gluon 
plasma\cite{RISCHKEandGYULASSY}.
For this reason we have been carrying out a program of lattice
simulations to determine this equation of state, or energy and pressure
as a function of temperature.  Our calculations include two degenerate
quark flavors.  However, the strange quark is neglected, and we work
at zero net baryon density.

While the approximations made in a lattice simulation are controllable
in principle, with presently available computing power these
approximations are severe.  In particular, effects of the nonzero
lattice spacing are large.  For example, there are big differences
between the continuum Stefan-Boltzmann law, which is presumably the
limit of the QCD energy at very high temperatures, and the lattice
version, obtained by summing over the Fourier modes with the free
particle action on the lattice.  Also, with the Kogut-Susskind quarks
that are usually used in high temperature QCD simulations, flavor
symmetry (isospin symmetry, more or less) is badly broken.  Instead
of having $N_f^2-1$ light pseudo-scalar particles at low temperature,
only one pion is an exact Goldstone boson corresponding to a symmetry
not broken by lattice artifacts.  The breaking of flavor symmetry
is expected to be proportional to $a^2$ ($a$ is the lattice spacing).
It can be reduced by modifying the action\cite{FLAVOR} or by
simply decreasing the lattice spacing. 

In this work we report on an extension of our equation of state
calculations to six time slices~\cite{NT6PRELIM}, which is 
a lattice spacing $a=1/(6T)$ instead of
the $a=1/(4T)$ used in our earlier work~\cite{NT4}.  
In addition to decreasing the
lattice spacing, we improve the extrapolation of our results to smaller
quark mass by fitting our free energy to a form with either the
theoretically predicted O(4) critical behavior or with the mean field
behavior that is expected when we are not very close to the critical point.

\section{Theory}

The methods for computing the energy and pressure are
standard\cite{karsch_coeff,Potvin90,Engels90}, and
we have discussed them in our earlier paper\cite{NT4}.
Here we summarize the equations necessary to make this paper
self contained.

The energy, pressure and interaction measure are defined by
\BEA
\varepsilon V &&=-{\PAR{\log Z}{(1/T)}} \EL
{\frac{p}{T}} &&={\PAR{\log Z}{V}} \EL
I &&= E - 3P = \frac{-T}{V} \PAR{\log(Z)}{\log(a)}
\la{epidef}
\EEA
The temperature and volume are determined by the lattice spacing $a$ and
the space and time dimensions of the lattice, $N_s$ and $N_t$.
\BEA
V = N_s^3 a^3,\EL
1/T = N_ta\ \ ,
\label{vandt}
\EEA
We use the $1\times 1$ plaquette (Wilson) action for the gauge fields, and
the conventional Kogut-Susskind quark action with two flavors
of quarks.
\BE
Z = \int [dU_{(n,\mu)}]
 \exp \{ \,\,(\gbeta) S_g +
(n_f/4)\, \Tr \log[am_q + \Dslash]\,\, \}\ \ ,
\la{defz}
\EE
\BE
S_g = {\frac{1}{3}}{\rm Re} \sum_{n,\mu<\nu}
\Tr{U_{\Box}{(n,\mu,\nu)}}\ \ ,
\la{gauge_action}
\EE

In a lattice simulation we compute derivatives of the partition
function.  From these we can either explicitly or implicitly
construct the partition function and from that the energy and pressure.
In this work we make use of these quantities:
\BE
\langle\Box\rangle = \frac{1}{2N_s^3N_t}
\PAR{\log(Z)}{6/g^2}
\label{plaq eq}
\EE
\BE
\langle\bar\psi\psi\rangle = \frac{1}{N_s^3N_t}
\PAR{\log(Z)}{am_q}
\label{pbp eq}
\EE
\BE
\langle\Box^2\rangle -\langle\Box\rangle^2 = \frac{1}{4N_s^3N_t}
\PARTWO{\log(Z)}{(6/g^2)}
\label{dplaq eq}
\EE
\BE
\langle\Box\bar\psi\psi\rangle - \langle\Box\rangle\langle\pbp\rangle =
\frac{1}{2N_s^3N_t} \PARTWOMIX{\log(Z)}{am_q}{6/g^2}
\label{dpbp eq}
\EE
The other second derivative of $\log(Z)$, 
$\partial^2{\log(Z)}/\partial(am_q)^2$, involves a disconnected piece
which is not calculated here.

For large systems the free energy is proportional to the volume, and
the pressure becomes just
\BE p = \frac{T}{V} \log(Z) \EE
Then derivatives of the free energy are just derivatives of the
pressure, and the pressure can be reconstructed by integrating
the free energy
\BE\la{PfromBETA_EQ}
{\frac{pV}{T}}(\gbeta,am_q) = 2N_s^3 N_t
\int_{\rm cold}^{\gbeta}
[ \LL {\Box} (6/g'^2,am_q)\RR - \LL {\Box} (6/g'^2,am_q)\RR_{\rm
sym}]d(6/g'^2)
\ \ ,
\la{int_plaq}
\EE
or
\BE\la{PfromMASS_EQ}
{\frac{pV}{T}}(\gbeta,am_q) = N_s^3 N_t
\int_{\rm cold}^{am_q}
 [\LL \pbp (\gbeta,m_q'a)\RR - \LL \pbp(\gbeta,m_q'a)\RR_{\rm sym}]
d(m_q'a)
\ \ ,
\la{int_pbp}
\EE
where the ``symmetric'' quantities subtract the divergent 
zero-temperature pressure.
The interaction measure can be found from simulations at a single
value of $(6/g^2,am_q)$ and the beta function, which tells how these
lattice couplings must be changed to change the lattice spacing.
\BE\la{I_EQ}
{\frac{I V }{T}} =
\,-\,2 N_s^3 N_t {\PAR{(\gbeta)}{\log a}}[ \LL \Box \RR-  \LL \Box
\RR_{\rm s
ym}]-
N_s^3 N_t {\PAR{(am_q)}{\log a}}[\LL \pbp \RR - \LL \pbp \RR_{\rm sym}]
\EE
In reference\cite{NT4} we obtained the nonperturbative beta function 
for couplings associated with the $N_t=4$ and 6 crossovers. To briefly 
summarize, zero temperature spectrum data from the literature
are combined in a fit which gives $am_\pi$ and $am_\rho$ as smooth functions
of $6/g^2$ and $am_q$.
The $\rho$ mass is used to define the lattice spacing.  In particular,
we somewhat arbitrarily set the $\rho$ mass to 770 MeV at all light
quark masses.
Lines of constant physics (\ie, $m_\pi/m_\rho={\rm constant}$),
along which the lattice spacing varies, are determined in the bare 
coupling space.
The two components of the beta function,
$ \partial(\gbeta)/(\log a)$ and $ \partial(am_q)/\partial(\log a)$,
tell how the input parameters change along these lines of constant
physics.


\section{ Simulations }

For the equation of state, we have carried out simulations using the 
hybrid molecular dynamics $R$~algorithm\cite{R_alg}.
The calculation requires both asymmetric ($12^3\times6$) and symmetric ($12^4$)
lattices as mentioned above. These are commonly referred to as hot and cold
lattices, respectively, and we will use this terminology.
This is somewhat misleading
since the system may be in the cold phase, \ie below the crossover, 
even on the hot lattices if the coupling is small enough. 
Each hot (cold) simulation is at least 1800 (800) time units long after 
equilibration. 
As a rule, fewer trajectories were required for the "cold" lattices
to achieve the same level of statistical accuracy because
the lattice volume was twice as large.
On the hot lattices near the crossover region the simulations were 
extended to more than 3000 units to overcome the larger fluctuations
associated with the transition.

\begin{figure}
  \vbox{
	\epsfxsize=6.0in \epsfbox[0 0 4096 4096]{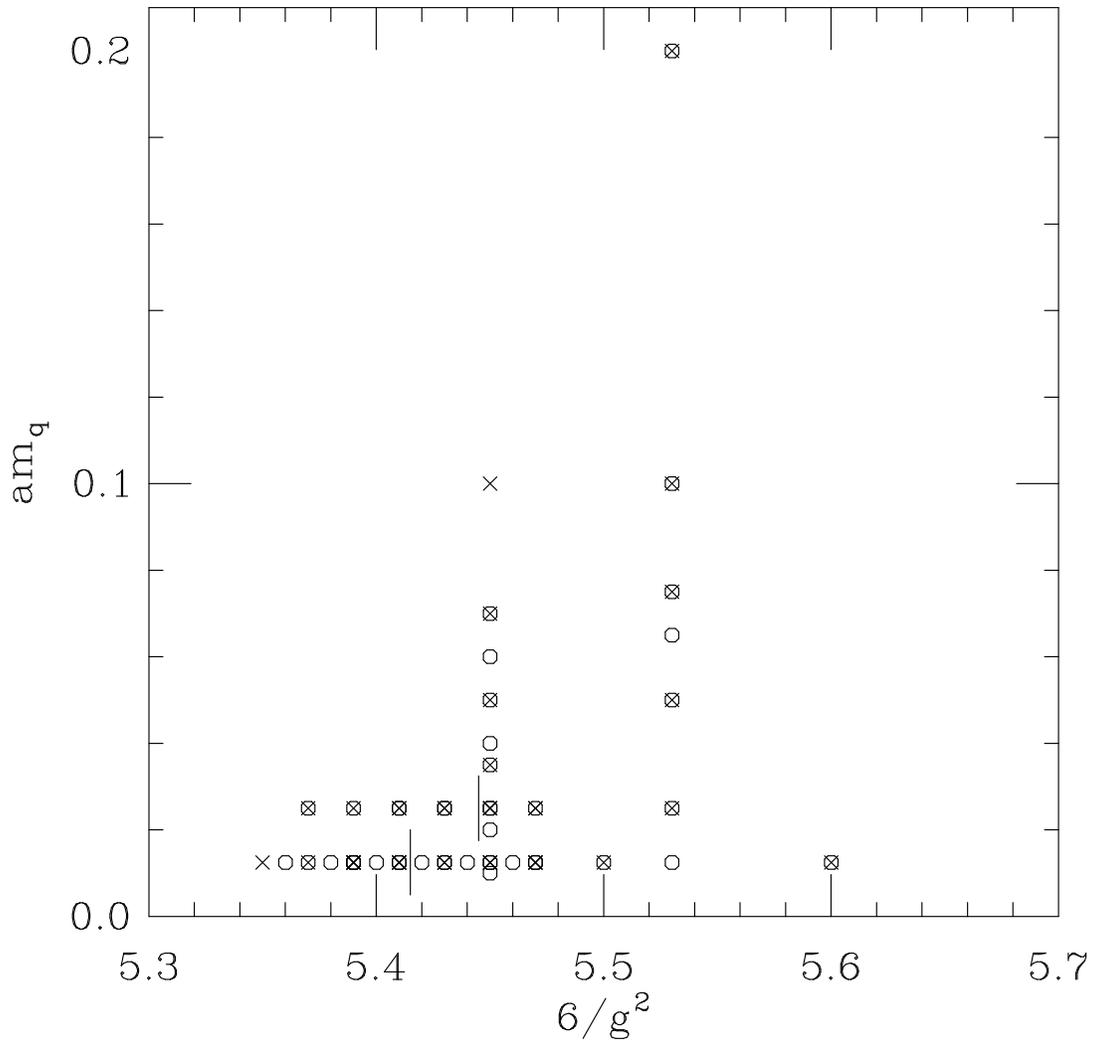}
  }
  \caption{Phase diagram for our simulations. The vertical solid lines indicate
	approximate locations of the crossover. Crosses indicate cold
	lattices, octagons hot lattices.
  }
  \label{runplot}
\end{figure}

The bare coupling and quark mass phase diagram corresponding to our simulations
is shown in Fig.~\ref{runplot}. The vertical bars denote 
the approximate location
of the finite temperature crossover for $am_q=0.0125$ and 0.025. 
Note, while increasing the coupling
$\gbeta$ is analogous to increasing the temperature, lowering the bare quark
mass $am_q$ also increases the temperature.
(This statement depends on the physical quantity used to define the
lattice spacing.  Again, we use $m_\rho$.)
Of course there are lines
in this bare coupling phase diagram on which $m_\pi/m_\rho$ is fixed 
and only the physical temperature of the system varies. 
In particular, the bottom of the graph,
where $m_\pi/m_\rho=0$, is
the line $am_q=0$ and corresponds approximately to the real world.
In our simulations the gauge coupling $\gbeta$ takes the values 
$5.35\leq\gbeta\leq5.6$ for quark mass $am_q=0.0125$ and 
$5.37\leq\gbeta\leq 5.53$ for $am_q=0.025$.
Then along the line $\gbeta=5.45$, the quark mass varies in the range 
$0.01\leq am_q\leq 0.1$ and at $\gbeta=5.53$, $0.0125\leq am_q\leq 0.2$. 
This range of couplings and masses corresponds roughly to physical
temperatures $125<T<250$ MeV (based on the $\rho$ mass\cite{NT4})
and mass ratios $0.3<m_{\pi}/m_\rho<0.7$.
In units of the pseudo-critical
temperature $T_{c}$, $m_\pi/T_{c}=1.94$ and 2.69 for $am_q=0.0125$ and
0.025, respectively. Past lattice simulations indicate that 
$T_c/m_\rho\approx0.2$\cite{NT12}, or $m_\pi/T_c\approx 0.9$ in
the continuum limit. Thus our simulations correspond to rather heavy
pions.

\begin{figure}
  \vbox{
	\epsfxsize=6.0in \epsfbox[0 0 4096 4096]{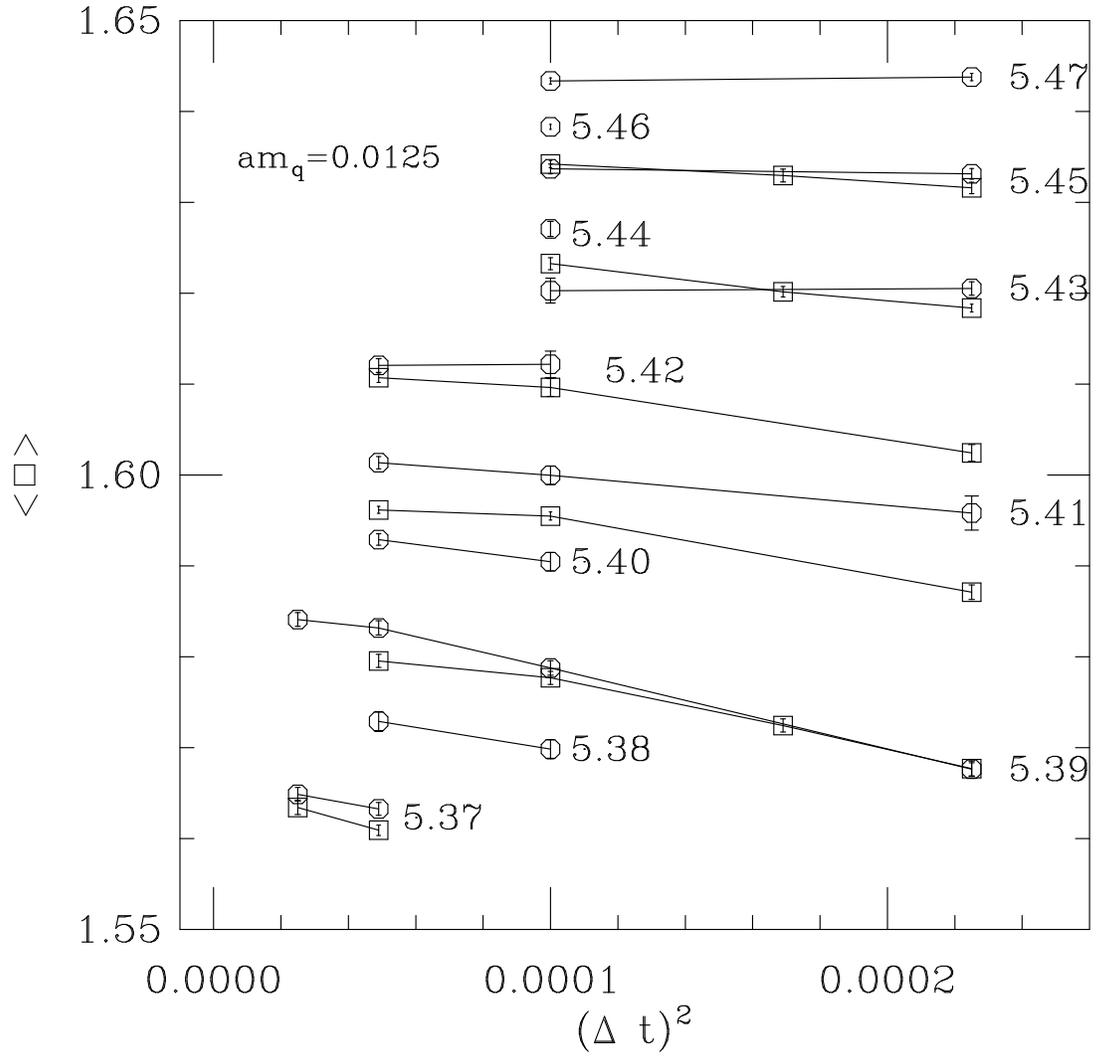}
  }
  \caption{ The plaquette expectation value as a function of
	the step size squared, $\dt^2$, and the gauge coupling. Results
	are shown for $am_q=0.0125$. Octagons denote hot 
	lattices ($\gbeta$ is given for each) and squares denote cold lattices.
  }
  \label{plqvsstep}
\end{figure}

The $R$~algorithm introduces lowest order errors in observables 
that are proportional
to $\dt^2$, where $\dt$ is the step size used to numerically integrate the
gauge field equations of motion through simulation time\cite{R_alg}. The
errors are in general different on hot and cold lattices, thus
multiple simulations at each value of $\gbeta$ and $am_q$ are required to
extrapolate observables to $\dt=0$\cite{NT4}. Of course, this greatly increases the
computational cost of the calculation since at each time step the force
term due to the quarks in the gauge field equations of motion requires
the inverse of the quark matrix, and as it happens, $\dt$ must be taken
relatively small ($\simeq am_q$) to be in the regime where the lowest
order error dominates. Thus many inversions of the quark matrix are 
required for each simulation. 
In Fig.~\ref{plqvsstep} we show example results for the 
plaquette ($\eplaq$) at $am_q=0.0125$. Evidently, on the cold lattices and 
for lower values of $\gbeta$, the effects are worse. The step size
errors are particularly troublesome for the plaquette since the
difference of the plaquette on hot and cold lattices is quite
small (even in the hot phase), as we will see in the next section.
At low temperature both $\epbp$ and $\eplaq$ approach their symmetric
values, making extraction of thermodynamic quantities in the hadronic
phase very difficult.

\section{Results and Analysis}

\begin{figure}
  \vbox{
	\epsfxsize=6.0in \epsfbox[0 0 4096 4096]{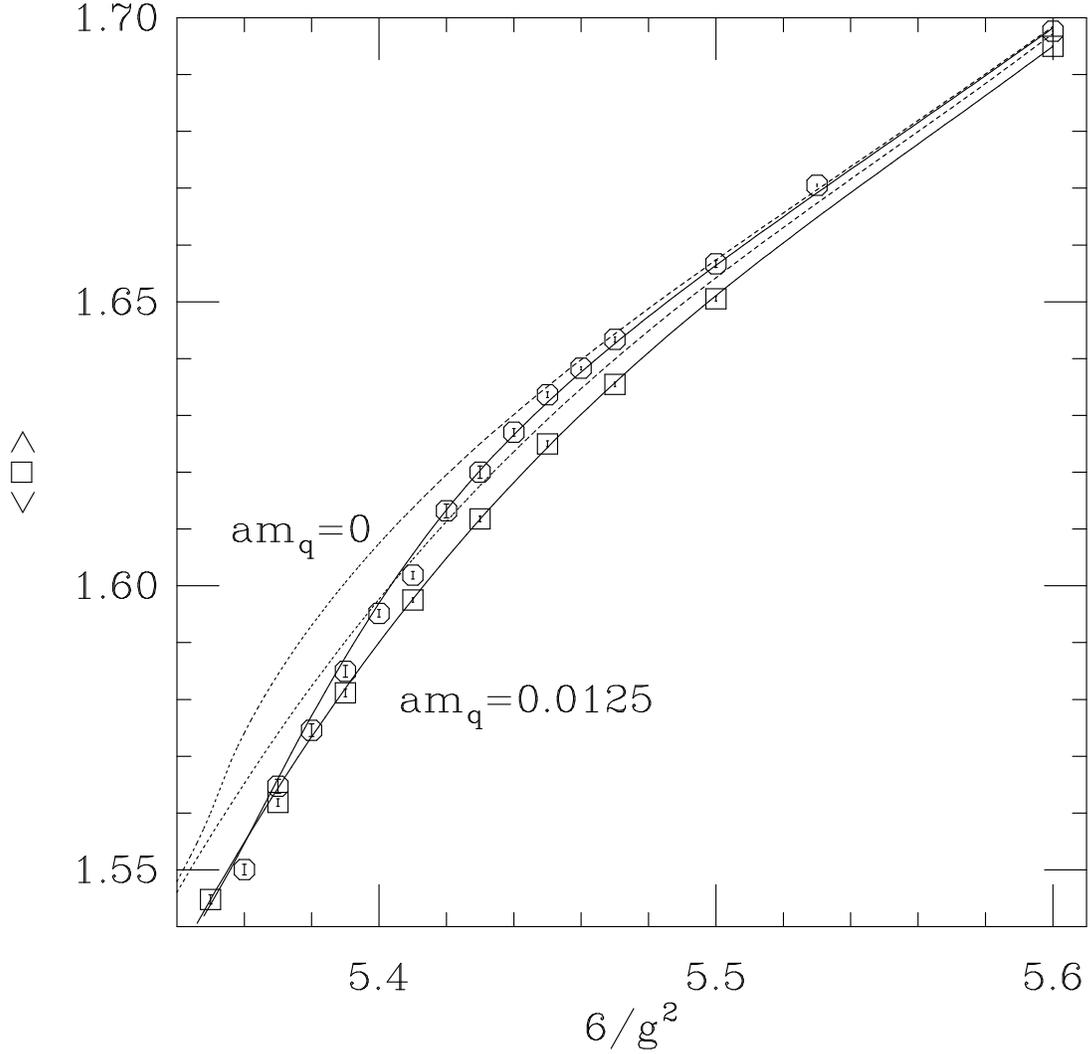}
  }
  \caption{ The plaquette expectation value as a function of
	the gauge coupling at $am_q=0.0125$.
	Octagons denote hot lattices, squares cold lattices. The
	area between the curves gives the pressure while the difference
	at each point is related to the interaction measure. The 
	lines are from a fit including the O(4) singular free energy 
	described in the text (dashed lines are an extrapolation to 
	$am_q=0$).
  }
  \label{plqvsbeta}
\end{figure}

The expectation values of $\eplaq$ and $\epbp$ are shown in 
Figs.~\ref{plqvsbeta} and~\ref{orderparam}. These values reflect the
step size extrapolations. In each figure, a small but noticeable
inflection point is observable in the expectation values on
the hot lattices at $\gbeta\approx 5.415(am_q=0.0125)$ and 5.445($am_q=0.025$). 
These couplings correspond to the pseudo-critical temperatures of the
crossover. The expectation values are smooth on the cold lattices. Note
that the values on the hot lattices approach their cold values at small
$\gbeta$ and begin to separate as $\gbeta$ (and thus the temperature) is increased. 
The area between the curves in Fig.~\ref{plqvsbeta} yields $p/T^4$.
At large $\gbeta$, the plaquette expectation values again approach each other
since $p/T^4\to {\rm constant}$ as $T\to\infty$, as expected 
from asymptotic freedom of QCD. 
The qualitative behavior of $\epbp$
is consistent with its expected role as an order parameter
for a spontaneously broken chiral symmetry.  A quantitative
analysis and extrapolation to zero-quark-mass is given later.

\begin{figure}
  \vbox{
	\epsfxsize=6.0in \epsfbox[0 0 4096 4096]{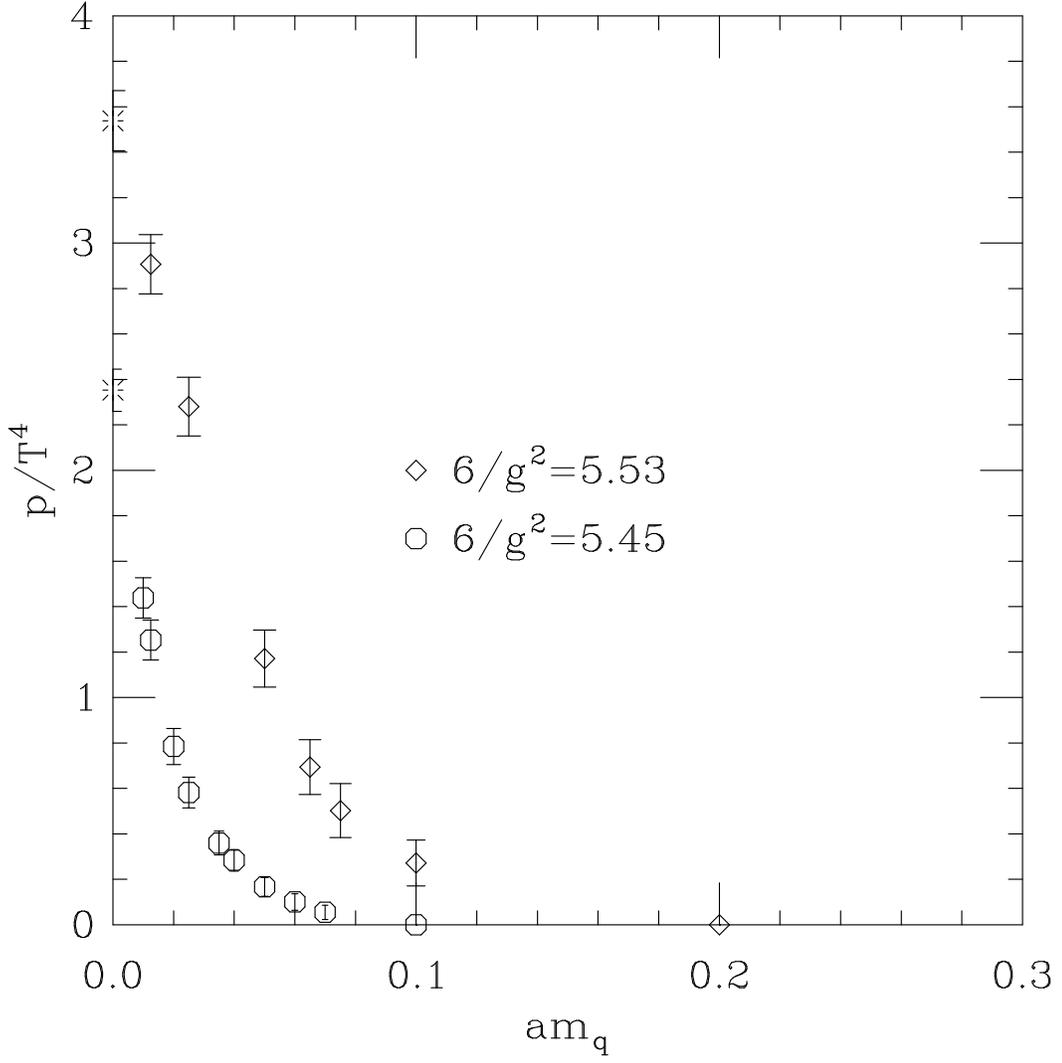}
  }
  \caption{ The pressure obtained from integrating $\epbp$ with
	respect to $am_q$ and constant $\gbeta$. The values at
	zero-quark-mass (bursts) are obtained by setting $\epbp=0$ on the
	hot lattices and extrapolating $\epbp$ to $am_q=0$ on the cold lattices.
  }
  \label{pvsmq}
\end{figure}

Next we turn to a discussion of the pressure. The integration of 
$\epbp$ with respect to $am_q$ at $\gbeta=5.45$ and 5.53
yields the pressure as a function of $am_q$ at fixed $6/g^2$,
shown in Fig.~\ref{pvsmq}. For large $am_q$ the system
is in the cold phase, and as $am_q$ decreases, the pressure smoothly increases.
To extrapolate the pressure from our smallest quark mass
to zero-quark-mass requires a corresponding extrapolation
of the cold and hot contributions to the integrand in (\ref{int_pbp}).
For the hot lattices at both values of $6/g^2$ the extrapolation
takes place in the hot phase, so for this contribution we assume
$\epbp = 0$ at $am_q = 0$.  For the cold
lattices we extrapolate using the fit summarized in Table~\ref{FIT TABLE}. 
A simple linear extrapolation of the hot lattice data for the
smallest two quark masses
gives a result within $1.5\sigma$ of zero at $6/g^2 = 5.45$
and $3.0\sigma$ of zero at $6/g^2 = 5.53$, perhaps due to
curvature effects or underestimated errors.  In any event
forcing a zero intercept has a negligible effect on the
pressure extrapolation.

\begin{figure}
  \vbox{
	\epsfxsize=6.0in \epsfbox[0 0 4096 4096]{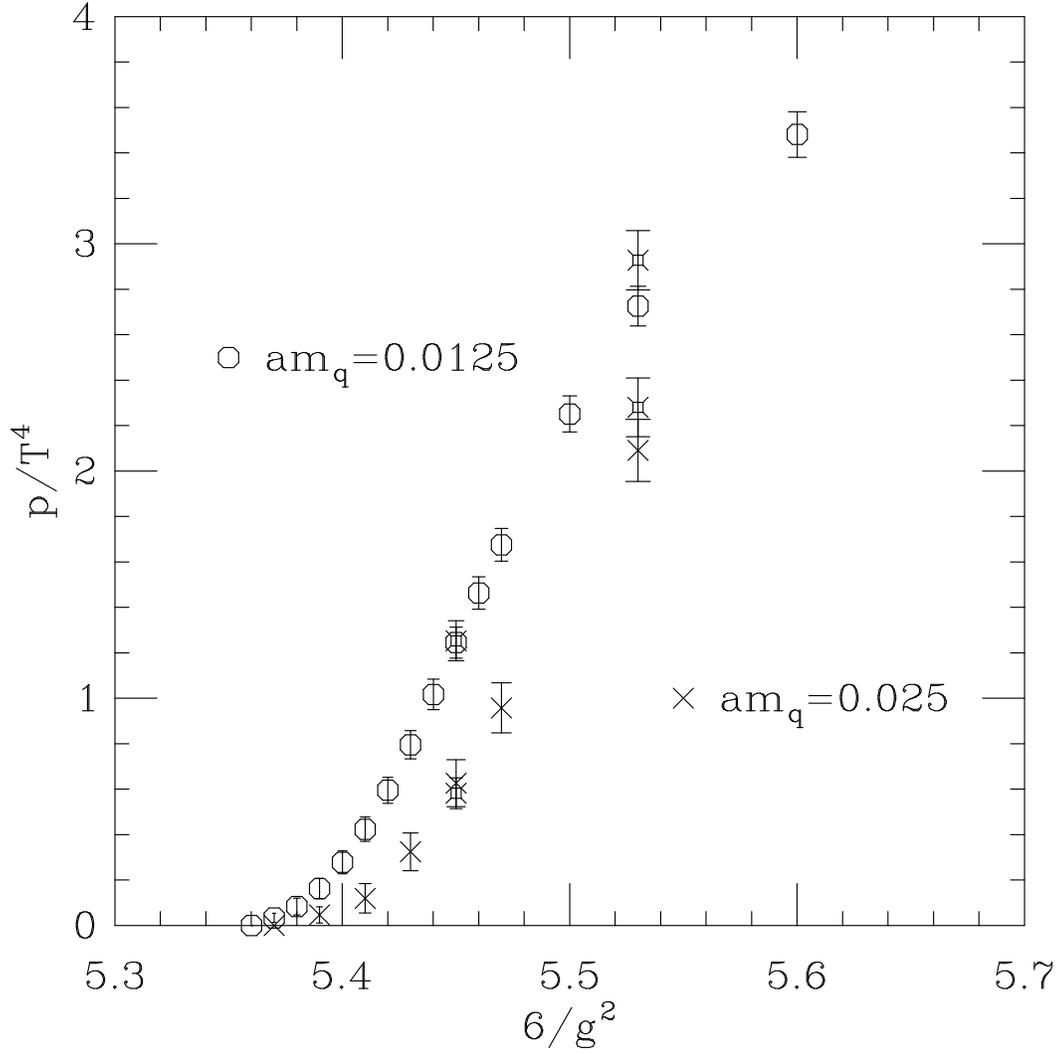}
  }
  \caption{ The pressure obtained from integrating $\eplaq$ with
	respect to $\gbeta$ at constant $am_q$. The values
	from the quark mass integrations (fancy squares) are also
	shown for comparison.
  }
  \label{pvsbeta}
\end{figure}

As mentioned above, the pressure as a function of $\gbeta$ 
at fixed quark mass is 
obtained by integrating $\eplaq$ with respect to $\gbeta$.
In Fig.~\ref{pvsbeta} we show the results
for $am_q=0.025$ and 0.0125. Again, the pressure rises smoothly through the
crossover. The curves are similar except for an overall shift in $\gbeta$,
the $am_q=0.0125$ curve beginning its rise sooner since the smaller quark mass
corresponds to higher temperature. 
Shown where they can be compared are the points obtained from the
the quark mass integration. The values from the two
different approaches agree, indicating
that the integration method works well for the volumes studied.

\begin{figure}
  \vbox{
	\epsfxsize=6.0in \epsfbox[0 0 4096 4096]{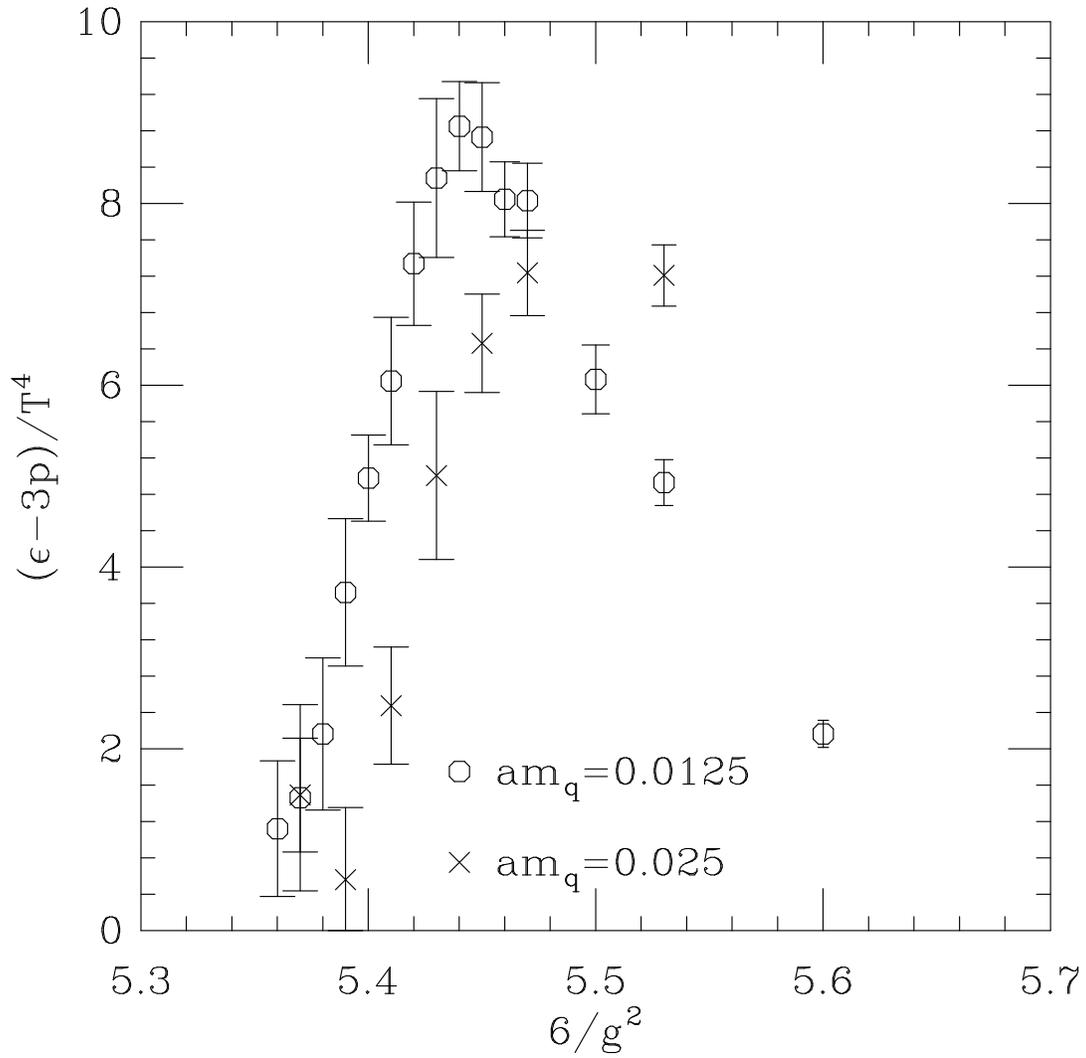}
  }
  \caption{ The interaction measure, $\varepsilon-3p$.
  }
  \label{imvsbeta}
\end{figure}

The interaction measure at each point is just the sum of the 
$T=0$ subtracted values
of the $\eplaq$ and $\epbp$ weighted by the coupling and quark mass components of
the $\beta$ function, respectively (Eq.~\ref{I_EQ}).
The results for $am_q=0.0125$ and 0.025
are shown in Fig.~\ref{imvsbeta}. At zero temperature the interaction measure
is zero since we have normalized $\varepsilon$ and $p$ to be zero at $T=0$. 
Through
the transition, we expect $I$ to increase rapidly if the energy density
increases rapidly, {\it e.g.} if the quarks and gluons deconfine, since the 
pressure must rise smoothly (even for a discontinuous
transition) to maintain mechanical equilibrium of the system. This 
behavior is seen in Fig.~\ref{imvsbeta}. Again, at higher temperature due
to asymptotic freedom,
we expect $I$ to decrease to zero as both the energy density and 
pressure asymptotically approach their Stefan-Boltzman values, and
$\varepsilon=3p$, the equation of state for a relativistic free gas.

\begin{figure}
  \vbox{
	\epsfxsize=6.0in \epsfbox[0 0 4096 4096]{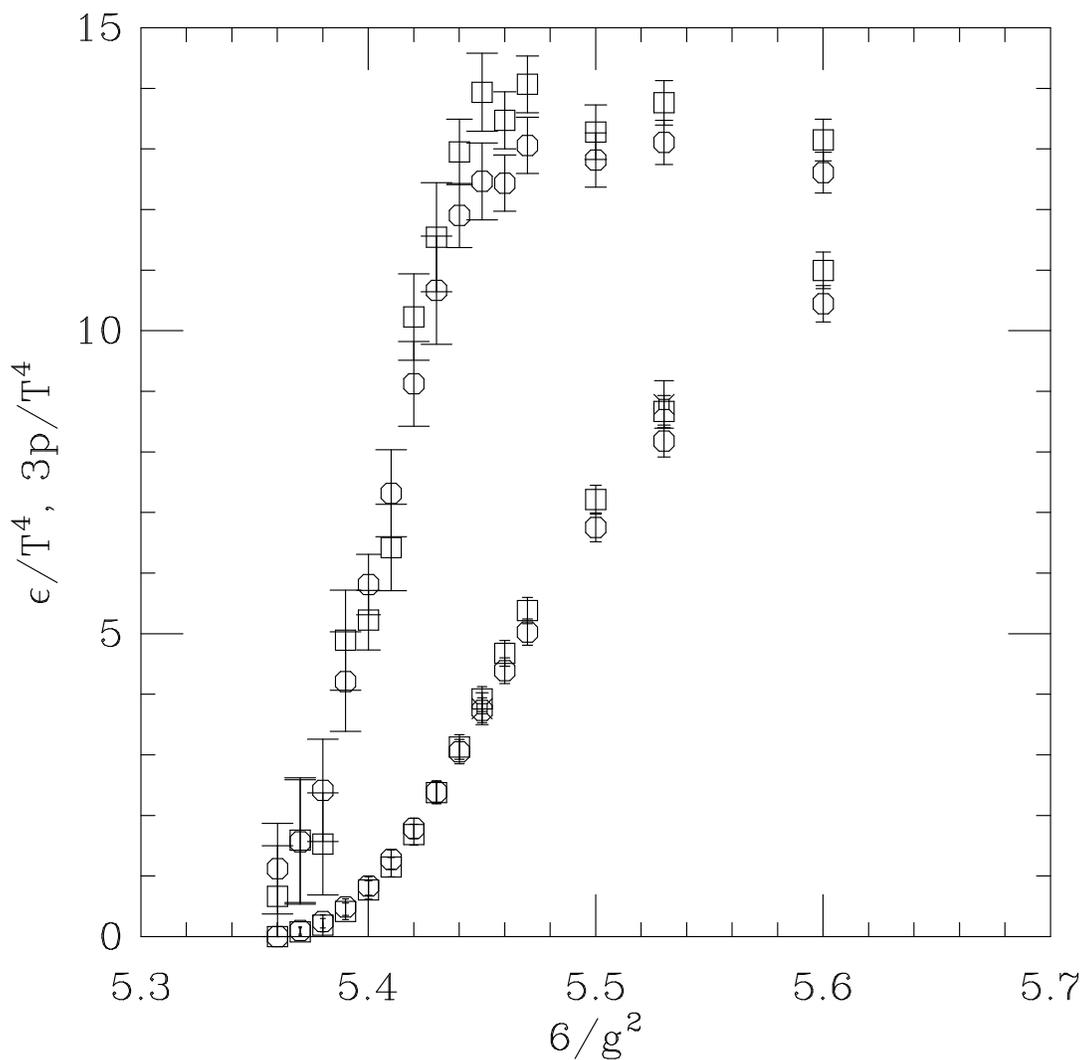}
  }
  \caption{ The energy density constructed from $I+3p$ (upper two curves). 
	Also shown is $3p$ (lower two curves) 
	and results with no step size extrapolations (squares).
	The difference is the step size systematic error in the equation
	of state. The data are for $am_q=0.0125$ only. The fancy squares 
	denote 3 times the pressure as calculated from 
	the integration of $\epbp$ with respect to $am_q$.
  }
  \label{syserror}
\end{figure}

The energy density constructed from $I$ and $p$ is shown in 
Fig.~\ref{syserror}.
Here we include $3p$ as well. We also plot $\varepsilon$ and $3p$,
using the values of the $\eplaq$ and
$\epbp$ at the smallest step size available at each point, without step size 
extrapolations.
The difference between the two gives an estimate of the step size systematic 
error in our final results. Recall that step size effects for larger masses 
were essentially eliminated by taking $\dt\ll am_q$.

\begin{figure}
   \vbox{\epsfxsize=6.0in \epsfbox[0 0 4096 4096]{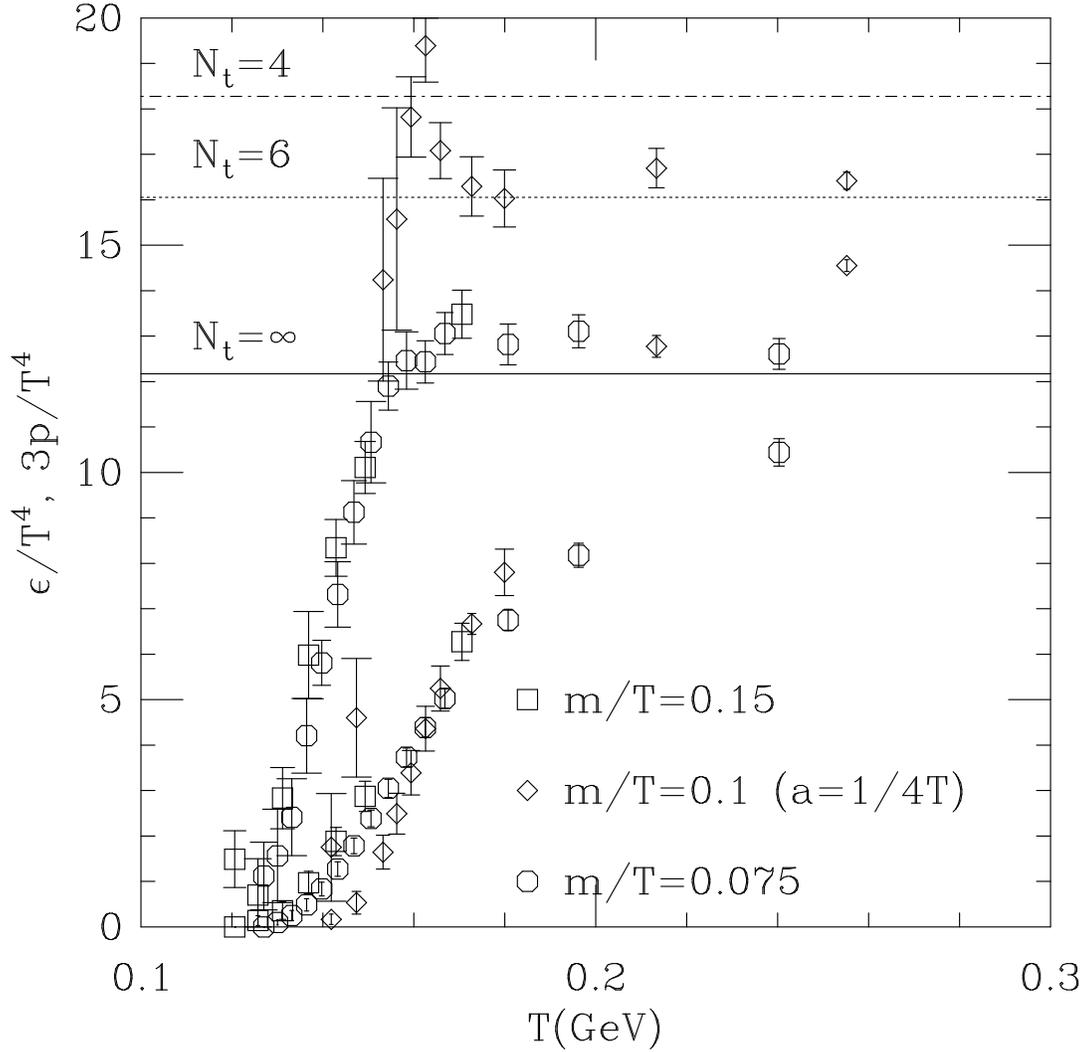} }
   \caption{ The equation of state along lines of constant $m_q/T$. The
	octagons denote $am_q=0.0125$ results, squares $am_q=0.025$.
	The diamonds indicate an earlier result on $N_t=4$ lattices. Horizontal
	lines correspond to Stefan-Boltzmann laws for $N_t=4$, 6, and the
	continuum. The energy density increases rapidly near the crossover 
	while the pressure (lower curve for each symbol) rises smoothly.
	}
   \label{eos}
\end{figure}

In Fig.~\ref{eos} we show the equation of state as a function of the
physical temperature. The quark mass dependence is largely removed
by this partial rescaling, $\{\gbeta(a), am_q(a)\}\to \{T(\gbeta,am_q),am_q\}$. 
Note that the physical quark mass, or more precisely $m_\pi/m_\rho$, varies
along each line of constant $am_q$ while $m_q/T=am_q N_t$ is held fixed.
From the figure, we see a large increase in the energy density as $T$ increases
through the pseudo-critical temperature $T_{c}\approx 140$ MeV 
($\gbeta_{c}$ was defined above). Just below $T_{c}$,
$\varepsilon/T^4\approx 5$ which is already substantial. Physically this means
the hadronic phase has non-negligible energy density, except that it is 
not clear what degrees of freedom are being excited since we have already
mentioned that $m_\pi/T_{c}$ is roughly 2 to 3. A similar situation exists
for the SU(3) pure gauge case where the ratio of the lightest glueball state 
to $T_c$ is greater than 5\cite{PUREGAUGE}. We also note that the energy
density of a relativistic gas of three light pions is insufficient to explain
the observed energy density in the hadronic phase
($\varepsilon/T^4=\pi^2/30\times 3\approx 1$). Similarly,
in SU(2) pure gauge theory it has been noted that a gas of 
non-interacting glueballs cannot account for the 
observed energy density below $T_c$\cite{SU2}.

In Fig.~\ref{eos} we also compare the $N_t=6$ equation of state with 
an earlier result on
$N_t=4$ lattices and with the continuum and lattice Stefan-Boltzmann laws. 
There is an apparent large finite
size effect which is expected from the free lattice theory. 
At high temperature the energy density has leveled off dramatically while
the pressure is still increasing at the largest value of $T$ that we simulated.
Because of asymptotic freedom, $\varepsilon$ and $3p$ should approach
the Stefan-Boltzmann result for the corresponding value of $N_t$. 
But, from Fig.~\ref{eos}, the approach to the free result is evidently quite 
slow.

\subsection{Extrapolation to $m_q=0$}

The above results pertain to unphysical values of the quark mass.
Indeed, we would like to obtain the equation of state along a line of
constant physics corresponding to the real (two flavor!) world. This
can be done by extrapolating to the chiral limit, $am_q\to 0$. To this end
we fit the derivatives of the free energy density to an appropriate 
function of $\gbeta$ and $am_q$ and then set $am_q=0$. The
fit serves the dual purpose of smoothing the data and allowing
a parameterization of the equation of state in terms of the bare quantities,
from which we can extract, \eg, the speed of sound. 

If the QCD high temperature phase transition with two flavors is
a second order transition at zero-quark-mass driven by the restoration
of chiral symmetry\cite{RNG_FOR_QCD}, then the critical part of the
free energy should have a universal form, up to the scale of the
gauge coupling and quark mass\cite{RNG_TEXTBOOKS}.
The free energy is the sum of
an analytic piece and a scaling piece:
\BE f = f_a( \gbeta, am_q ) + f_s(t,h) \EE
where $t = (T-T_c)/T_0$ and $h=H/H_0$.  $T_0$ and $H_0$ are 
conventionally determined by requiring
$\epbp(t=0,h)=h^{1/\delta}$ and $\epbp(t<0,h=0)=(-t)^\beta$.
(In the language of spin models, $\epbp$ is the magnetization and
$\eplaq$ is the energy.)
From invariance under a length rescaling by a factor $b$,
the critical part of the free energy should have the property:
\BE\la{scaling_eq} f_s(t,h) = b^{-d} f_s(b^{y_t}t,b^{y_h}h) \ \ .\EE
This implies that the magnetization near the critical point
is determined by a universal scaling function, conventionally written as:
\BE\la{f_scaling_eq} \frac{M}{h^{1/\delta}} = f(t/h^{1/\beta\delta}) =
f(x) \\ .\EE
The normalization conditions on $t$ and $h$ then require that $f(0)=1$
and $f(x) \rightarrow (-x)^\beta$ as $x \rightarrow -\infty$.
This condition, along with the known values of $y_h$ and
$y_t$\cite{O4_EXPONENTS}, has been used to compare the behavior in $\epbp$
with that expected from O(4) symmetry\cite{O4_PREVIOUS}.  Here we
wish to use this theoretical input to guide extrapolation of
the free energy to physical light quark mass, which is essentially the
same as $am_q=0$.
We fit to both a scaling form for O(4) in three dimensions and to
the form for mean field theory.  We expect the mean field form
to be a good approximation when the system is not very near the critical
point, and the difference between these two forms gives an idea of
the systematic errors in this approach.

Since we calculate both $\epbp$ and $\eplaq$ (and their derivatives),
we would like to treat them equally in fitting the free energy.
Therefore we use a formulation of the scaling free energy which
handles the energy and magnetization symmetrically.  This has
been discussed in Ref.\cite{ON_SCALING}, so we just summarize
it here.

The scaling ansatz, Eq.~\ref{scaling_eq},
tells us that if we specify the singular free energy on
the unit circle in the $t,h$ plane, we have specified it for all $t,h$.
Thus the scaling part of the free energy density can be written
\BE f_s(t,h) =  b(t,h)^{-d}g(\theta(t,h)) \EE
where $b$ is the solution to
\BE\la{unit_eq} \LP b^{y_t} t \RP^2 + \LP b^{y_h} h \RP^2 = 1 \EE
and
\BE \theta(t,h) = \tan^{-1}(b^{y_h} h, b^{y_t} t) \ \ .\EE
Here $g(\theta)$ is a universal function.
In Ref.\cite{ON_SCALING} $g(\theta)$ for O(4) is determined approximately by
Monte Carlo simulation of the O(4) spin model.
For this formulation it is convenient to modify the conventional
normalizations of $t$ and $h$, and use $t=(T-T_c)/T_g$ and $h=H/H_g$,
where $T_g$ and $H_g$ are chosen to fix $g(\pi/2)
= y_h/d$ and $g^\prime(\pi) = -1$.
These are related to the conventional normalizations by 
$H_0=H_g^{\delta+1}$ and $T_0=T_g H_g^{1/\beta}$.

From simulation of the O(4) spin model an approximate
$g(\theta)$ for O(4)\cite{ON_SCALING} has been obtained.
For the mean field case, 
the scaling function
can be obtained from a numerical re-parameterization of the mean field
magnetic equation of state\cite{RNG_TEXTBOOKS}
\BE h/M^3 = 1 + t/M^2 \ \ .\EE

\begin{table}
\begin{tabular}{cccclc}
\hline
\hline
fit & $6/g^2_c$& $T_0$ & $H_0$ & $f_{ns}$ & $\chi^2$/dof\\
\hline
O(4) & 5.353 & 0.522 & 1.51
&$3.339\Delta\beta + 2.30\Delta\beta^2 + 0.98\Delta\beta^3 $ &\\ 
&&&&$+am_q^2(0.41 + 4.3\Delta\beta -8.3\Delta\beta^2)$
& 990/91 \\
MF & 5.381 & 0.546 & 1.22
&$3.221\Delta\beta + 0.505\Delta\beta^2 -0.221\Delta\beta^3 $  &\\
&&&&$+am_q^2(0.63 + 0.7\Delta\beta -0.6\Delta\beta^2)$
& 933/91 \\
cold & & & & $0.132\Delta\beta+
2.02\Delta\beta^2+
3.195\Delta\beta^3+
-1.23\Delta\beta^4$ & 130/52\\
&&&&$+am_q(
1.04\Delta\beta+
0.723\Delta\beta^2+
5.5\Delta\beta^3 -45\Delta\beta^4)$&\\
&&&&$+am_q^2(
-1.34\Delta\beta
-12\Delta\beta^2+
215\Delta\beta^3+
2.2\Delta\beta^4)$&\\
\hline
\hline
\end{tabular}
\caption{ Fit summary table. $f_{ns}$ is the non-singular free energy up
to an overall constant. Note $\Delta\beta\equiv 6/g^2-5.4$.}
\label{FIT TABLE}
\end{table}
%

Fig.~\ref{orderparam} shows $\epbp$ calculated from fits to the
mean field and O(4) scaling functions, plus polynomials in 
$am_q$ and $6/g^2$. We also include the pure polynomial fit to the 
cold data, and both hot and cold data are shown for comparison. The fits
are summarized in Table~\ref{FIT TABLE}. $\chi^2$
per degree of freedom is poor for all of the fits. For the hot data,
the mean field and O(4) cases each have $\chi^2/{\rm dof}\approx 1000/91$
while the cold data has $\chi^2/{\rm dof}\approx 130/52$. If we fit to
the data without step size extrapolations, the results improve somewhat,
$\chi^2/{\rm dof}\approx 650/91$ for the hot data and 117/56 for the cold. 
A fit to only $\epbp$ over the same range gives $\chi^2/{\rm dof}\approx 49/18$.
Despite the high $\chi^2$, examination of Fig.~\ref{orderparam}
shows that the data are actually reproduced quite well by the fits. This
is remarkable given the large range of $am_q$ and $\gbeta$ spanned by the fits.
Moreover, the fits to mean field and O(4) scaling functions work equally
well; our data cannot distinguish between the two scaling behaviors.
However, it is interesting to note that the respective extrapolations 
to $m_q=0$ are quite different, and give critical
temperatures $T_c\approx 140$ and 150 MeV for O(4) and mean field respectively.
The above indicates that the
present lattice simulations may still be too far from the
scaling region and smaller quark masses are required to see the true
scaling behavior. The correlation length in lattice units as given by the
inverse pion mass (the pion and sigma are degenerate at the critical point)
of the present simulations is only 2-3 which is less than the 
temporal extent of the lattice, so true dimensional reduction,
which must occur for universality arguments to hold,
has not been achieved.

At $am_q=0$, $\partial(am_q)/\partial(\log a)=0$ so the
interaction measure is determined solely from the plaquette.
However, most of our information about the critical behavior comes
from $\epbp$, since the contribution of the scaling part of the free
energy to the plaquette is small compared with the analytic part.
Therefore in a crude approximation our procedure is using information
about $\epbp$ to help determine the free energy, which in turn yields
the plaquette as $am_q \rightarrow 0$.

\begin{figure}
  \vbox{ \epsfxsize=6.0in \epsfbox[0 0 4096 4096]{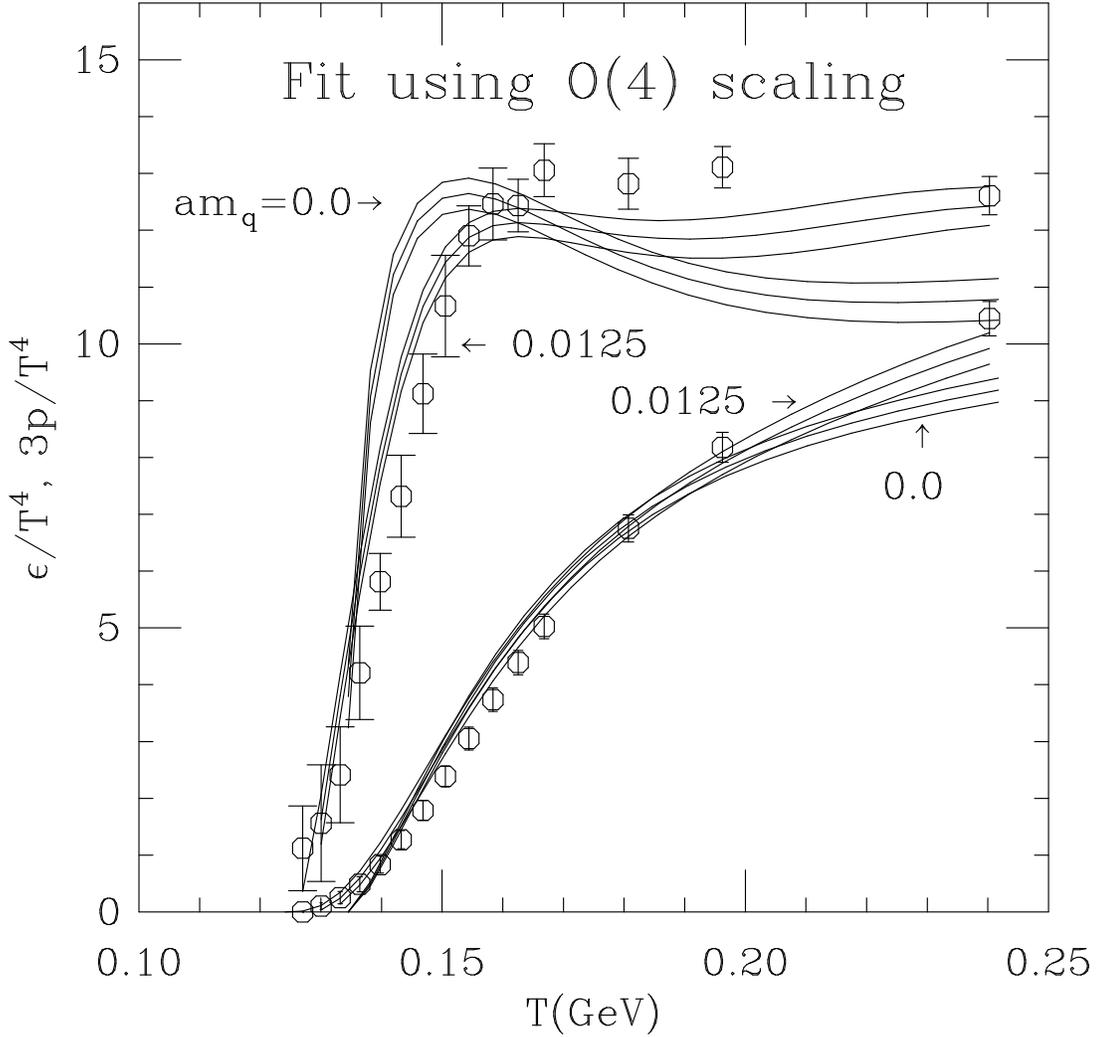} }
  \caption{ The $am_q=0$ equation of state from a fit to the data that includes
	the O(4) universal scaling function. Also shown is the data and
	the fit at $am_q=0.0125$ for comparison. Again, there is only a weak
	mass dependence. The ``bump" just after the transition is likely an
	artifact of the extrapolation. The solid lines correspond to the 
        central value and a one standard deviation 
	above and below this result (statistical errors only).}
  \label{eosfit}
\end{figure}

An extrapolation of the equation of state to $m_q=0$ is shown in 
Fig.~\ref{eosfit}. It
is compared to the $am_q=0.0125$ result which reproduces the data 
reasonably well. The appearance of the bump in the energy density just after
the transition is probably an artifact of the extrapolation (at $m_q=0$, 
the corresponding region of $\gbeta$ lies below the values of the coupling 
where we have done simulations). From Fig.~\ref{eosfit} we again see a 
weak dependence on the quark mass, which gives us some reassurance in the
extrapolation. The plus and minus one standard deviations shown in 
Fig.~\ref{eosfit} are calculated in the following way. First, 
we do a covariant fit 
to $\eplaq$, $\epbp$, and their derivatives with respect to $\gbeta$
(Eqs.~\ref{plaq eq}-\ref{dpbp eq}). From the fit we
obtain a set of parameters and the covariances of these parameters 
which map onto a multi-dimensional Gaussian probability distribution 
for the parameters. We then generate many parameter sets with this 
distribution and calculate the equation of state for each one. The
standard deviation of the mean of this set is shown in the figure.

\begin{figure}
  \vbox{ \epsfxsize=6.0in \epsfbox[0 0 4096 4096]{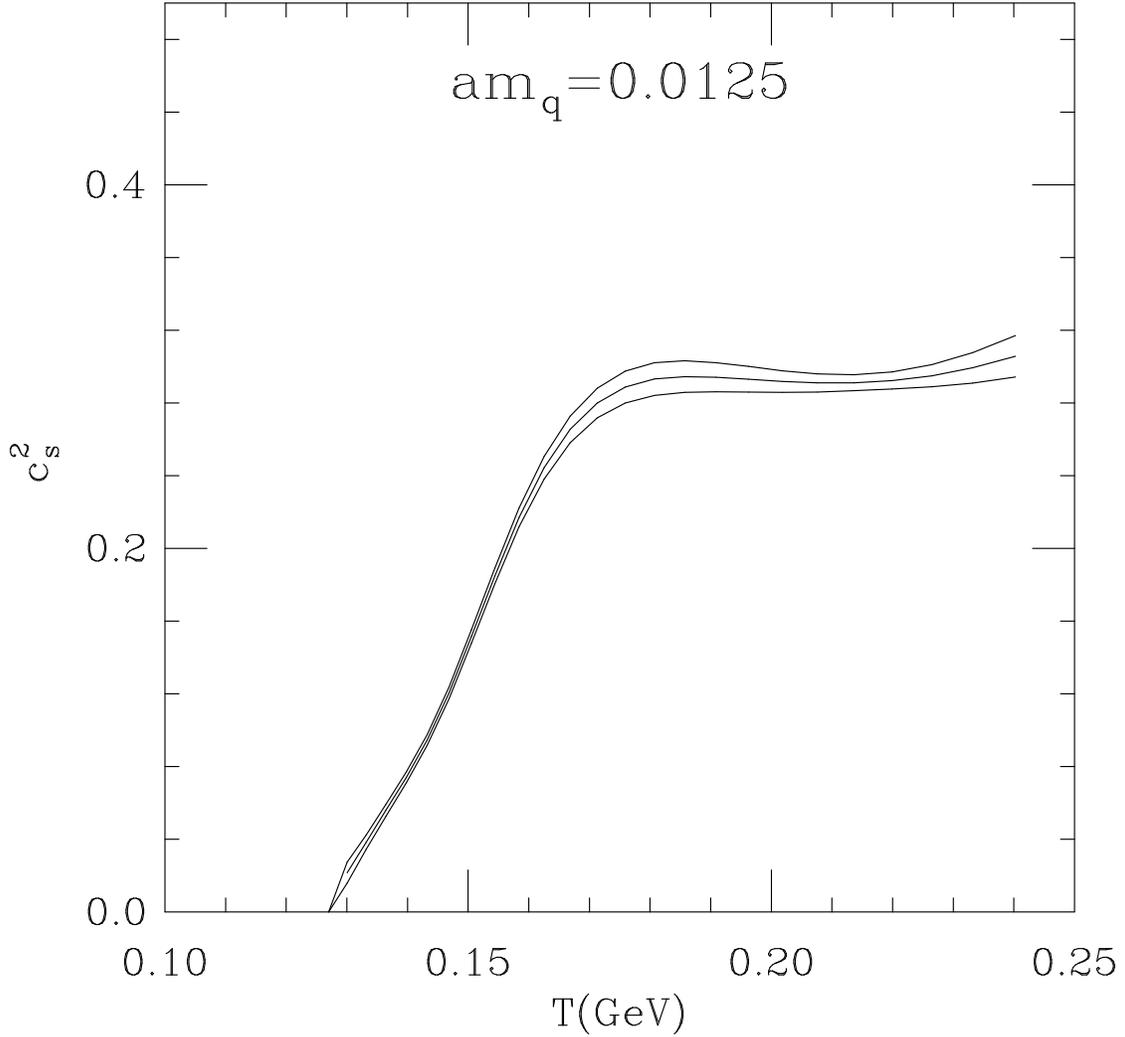} }
  \caption{ The speed of sound squared from a fit to the data that includes
	the O(4) universal scaling function. $c_s^2$ rises rapidly in the
	crossover region and then approaches the free gas result, 1/3.
	The low temperature result is probably not accurate (see text).  
	The solid lines correspond to the central value and a one standard 
	deviation above and below this result (statistical errors only).}
  \label{sound}
\end{figure}

In Fig.~\ref{sound} we show the 
speed of sound squared calculated from the O(4) fit. It rises rapidly through
the transition region and then levels off near the free value of 1/3.
This indicates that the system is weakly interacting in this region. This should
be contrasted with $\varepsilon - 3p$ just after the crossover which indicates 
significant interaction effects. Indeed, the couplings in the region are of order one,
and we have already seen that neither the energy density nor pressure
is approaching its perturbative value. The low temperature part of 
the curve is probably not accurate. The derivatives of the energy 
density and pressure are poorly known in this region since the difference
of $\eplaq$ and $\epbp$ from their cold lattice values is nearly zero.
We have already
mentioned that $m_\pi/T_{c}$ was rather high in our simulations, so
it is not surprising that the expected dynamics of a dilute gas of 
relativistic pions is not observable.
In that case, we expect the hadron gas below the transition to have a 
nonzero speed of sound, which then dips down at the transition. 
The statistical errors for the speed of sound were calculated in the same
manner described above for the equation of state.

\section{Conclusions}

In this work we have pushed our calculation of the equation of
state for QCD including dynamical quarks to smaller lattice spacing.
We have also developed techniques for using theoretical expectations for
the scaling behavior to extrapolate numerical results to the physical quark
mass.  
The points where we ran were chosen to explore the equation of
state over a range of temperature rather than to work very close
to the crossover.  Therefore, we can say little about the
nature of the transition or crossover from this work.  Still,
our results are consistent with the standard picture of a
second order phase transition
at zero-quark-mass and a sharp crossover for small but nonzero masses.
%

We remark, however, that recent simulations with two flavors of
Kogut-Susskind quarks on lattices with $N_t = 4$ and at smaller quark
masses than used in this study have revealed significant finite size
effects that have cast doubt on earlier promising demonstrations of
critical scaling\cite{O4SCALING}. Other recent simulations with two flavors of
Wilson quarks and an improved gauge action to reduce cutoff effects
also show promising agreement with O(4) scaling, but a thorough
investigation of finite-size effects remains to be done\cite{IWASAKI}. So for
the moment the question of the order and universality class of the
transition remains open, and the validity of our extrapolation to zero
quark mass remains to be established.  At the very least, one may
expect that the recently reported finite volume corrections lead to a
greater sharpening of the crossover in energy density and speed of
sound as the quark mass is decreased.  We have seen that small changes
in the extrapolation have a large effect on some but not all
extrapolated values: {\it e.g.}  O(4) and mean-field extrapolations
give zero-quark-mass critical temperatures that differ by 10 MeV.  On
the other hand, expressed as a function of temperature in physical
units, the energy density and pressure away from the crossover show
little dependence on quark mass, even in the zero-quark-mass limit.

The methods used here require a subtraction of the zero temperature
plaquette and $\epbp$.  As $N_t$ is increased, the plaquette subtraction
rapidly becomes more difficult, since the fractional difference in plaquette
between the hot and cold lattices decreases as $N_t^{-4}$.
This suggests that, as for many other quantities, an improved action
which allows use of a larger lattice spacing, or smaller $N_t$, will be
important for further progress.  Results for pure gauge theory and for
four-flavor QCD have been reported by the Bielefeld
group\cite{BIELEFELD_IMPROVED}.  Another important problem for future
studies is remedying the breaking of flavor symmetry, so that the low
temperature phase that is simulated really has three light pions.  For
a start in this direction see Ref.\cite{FLAVOR}.

\vskip 0.25in \centerline{Acknowledgments}
This work was supported by NSF grants
NSF-PHY93-09458, 
NSF--PHY96--01227, 
NSF--PHY91--16964, 
DOE contracts
DE-AC02--76CH--0016, 
DE-AC02--86ER--40253, 
DE-FG03--95ER--40906, 
DE-FG05-85ER250000, 
DE-FG05-92ER40742, 
and DOE grants
DE-2FG02--91ER--40628, 
DE-FG02--91ER--40661. 
Calculations were carried out on the following: 
the Intel Paragon at the San Diego Supercomputer Center, 
the Intel Paragon at Indiana University,
the IBM SP2 at the Cornell Theory Center,
the IBM SP2 at the University of Utah, and
the workstation cluster at SCRI.

\begin{figure}
\vbox{ \epsfxsize=6.0in \epsfbox[0 0 4096 4096]{orderparam.ps} }
  \caption{ $\epbp$ as a function of $\gbeta$ and $am_q$. The difference
	in the hot (octagons) and cold (squares) values is related to
	the interaction measure. $\epbp$ on the hot lattices also serves
	as an order parameter for the system. Lines depict fits to
	O(4) and mean field singular forms of the free energy plus
	analytic terms to $\eplaq$, $\epbp$, and their derivatives 
	(Eqs.~\ref{plaq eq}-\ref{dpbp eq}). Only points with $am_q\le 0.025$
	were included in the fits.
	The difference between O(4) (dotted line) and mean field forms
	(dashed line) is not discernible 
	at the quark masses where simulations were run. 
	Extrapolations to $m_q=0$, however, give
	different critical couplings. Solid lines correspond to polynomial 
	fits to the cold data and the corresponding extrapolation to $am_q = 0$.
  }
  \label{orderparam}
\end{figure}

\end{document}